\documentclass[12pt]{article}
\usepackage{mathspic, amssymb}
\begin{document}

\title{"Plane" electromagnetic wave\\
in spatially flat Friedman universe}
\author{Bernard Jancewicz\\
Institute of Theoretical Physics, University of Wroc{\l}aw, \\
pl. Maksa Borna 9, PL-50-204 Wroc{\l}aw, Poland\\
{\sf email: bjan@ift.uni.wroc.pl}} \maketitle

\begin{abstract}
The electromagnetic theory is, to a large extend, metric
independent. Before the metric is introduced, it is called premetric
electrodynamics. Metric enters the constitutive relation. We
consider this relation for the Friedman model of an expanding
Universe and find that the magnitudes of the field quantities depend
on the scale factor. This factor, however, does not enter the permeability
and permittivity of the vacuum. A spatially uniform electromagnetic field 
is obtained for spatially flat metric. Then plane electromagnetic wave is
found with uniform field playing the role of amplitudes. It turns out that 
the magnitudes of the frequency and the wave covector depend on the scale 
factor determining the redshift, but the phase velocity of the wave is constant.
\end{abstract}

\section{Introduction}

Solutions to Maxwell equations in the Friedman universe have been
presented in many papers \cite{schrod1}--\cite{haghi}.  They were using
antisymmetrc tensor $F^{\mu \nu}$ of the electromagnetic field, uniting the
fields $E$ and $H$. Two Maxwell equations were written only for $F$ field
with the use of space-time metric. The solutions (in analogy to quantum mechanics) 
were searched with a definite angular momentum, therefore the spherical harmonics have
been used. There were no attempts to look for uniform fields.

We are using another approach to electromagnetism with the aid of differential forms.
A way of presenting electrodynamics based on a broad use of differential
forms has been proposed in the last decades, see Refs.
\cite{Grauert}--\cite{Jancew1}. Within this approach, a question was
discussed whether electrodynamics can be formulated in such a way
that the metric of space-time doesn't enter the fundamental laws.
This framework is called {\it premetric electrodynamics}, see the
papers \cite{Jancew1}--\cite{Oziew}. Related developments can be
also found in the books \cite{Lindell},\cite{Russer}. A crowning
achievement of this approach is the book by Hehl and Obukhov
\cite{Hehl} in which classical electrodynamics is introduced
deductively, i.e. in the form of axioms: conservation of electric
charge, magnetic flux and energy-momentum. This approach uses
two electromagnetic field exterior forms, namely field strength 
$F$ and excitation $G$ and two separate Maxwell equations for them. 
At the end, the metric is introduced by the constitutive relations 
between $F$ anf $G$.

The splitting of $F$ and $G$ onto electric and magnetic parts depends on the observer
or, equivalently, on the coordinate system. Once the space-time coordinates 
$(t,x^1,x^2,x^3)$ are chosen, the four-dimensional two-form $F$ is expressed 
by electric field strength $E$ and magnetic induction $B$:
\begin{equation} F=E\wedge \underline{d}t +B, \label{faradsplit}
\end{equation}
and the four-dimensional two-form G -- by magnetic field strength $H$
and electric induction $D$:
\begin{equation} G=-H\wedge \underline{d}t +D, \label{gaussplit}
\end{equation}
where $\underline{d}$ is the exterior derivative.\footnote{$E,H$ are one-forms,
$B,D$ are two-forms.} After introducing
the three-form of the electric current density
\begin{equation} J=-j\wedge \underline{d}t +\rho \label{cursplit}
\end{equation}
the differential Maxwell equations are written as two equations
\begin{equation}\underline{d}\wedge F=0, \label{flux}\end{equation}
\begin{equation} \underline{d}\wedge G=J. \label{gauss}\end{equation}

Up to now all considerations are generally covariant and
metric-free. They are valid in flat Minkowskian as well as in
curved pseudo-Riemannian space-time. Therefore, these Maxwell's
equations represent the optimal formulation of classical
electrodynamics.

The electromagnetic theory at some moment incorporates the metric of
a flat or curved space-time via the constitutive relation between
the excitation and the field strength. In the present paper we
consider it in the form
\begin{equation} G=\lambda _0\star F, \label{constrim}\end{equation}
where the Hodge star \,$\star $\, is defined by the space-time
metric and $\lambda _0=\sqrt{\varepsilon _0/\mu_0}$ is the vacuum
admittance, the same as in the standard Maxwell-Lorentz
electrodynamics in the vacuum. It is our purpose to consider electromagnetic
field in a space-time of general relativity applied to cosmology. We assume that
the electromagnetic field is weak, which means that its energy momentum tensor
does not influence the gravitational field.

\medskip The simplest cosmological model is the Friedman solution \cite{demian} 
of the Einstein equations without cosmological copnstant with the following square
of the space-time distance expressed in so called {\it comoving frame} in which galaxies
have permanent positions $(x^1,x^2,x^3)$:\footnote{Here $dt,\, dr$ and so on denote the ordinary
differentials.}
\begin{equation}
ds^2=c^2dt^2-a(t)^2\left[ \frac{dr^2}{1-Kr^2}+r^2(d\theta ^2+\sin
^2\theta d\phi ^2)\right] ,\label{fried1}\end{equation} where $a(t)$
is the scale factor, $r^2=(x^1)^2+(x^2)^2+(x^3)^2$ and $\theta , \phi $ are
angles of spherical coordinates. The constant $c$ is the dimensional coefficient relating
units of time and space coordinates. Whether $c$ is velocity of light in this universe, will
be checked in Section 5. Since our aim is to find
spatially uniform fields, we choose the spatially flat universe with $K=0$
\begin{equation} ds^2=c^2dt^2-a(t)^2\left[ dr^2+r^2(d\theta ^2+\sin
^2\theta d\phi ^2)\right] ,\end{equation}
that is
\begin{equation} ds^2=c^2dt^2-a(t)^2\left[ (dx^1)^2+(dx^2)^2+(dx^3)^2\right]
=c^2dt^2-a(t)^2d{\bf r}^2. \label{fried2}\end{equation}

\medskip The scalar product determined by (\ref{fried2}) is used in Section 2
to define magnitudes of the electromagnetic field quantities, which
we consider to be measurable quantities. This notion was not yet
expressed in the literature. The named magnitudes turn out to depend
on the scale factor.

We exploit constitutive relation in Section 3 to determine the
permittivity and permeability of the vacuum. They
are not functions of time; their product is constant and related to
$c$ through $c=(\varepsilon _0\mu_0)^{-1/2}$.

Generally, a wave is a product of the amplitude and a periodic function
$\psi $ of the phase. In fundamental physics courses, the whole 
spatially-temporal dependence is present in $\psi $, hence the
amplitude is constant. When the metric is
time-dependent, the amplitude may depend on time, hence 
we look for an electromagnetic wave in two steps.
In Section 4, we consider an electromagnetic field $\tilde{F},\tilde{G}$ which could be
a counterpart of the static uniform field possible in Minkowski space-time.
We can not expect the field to be
static, but we choose it to be uniform, that is, independent of the spatial
coordinates. We find such solutions to homogeneous Maxwell equations i.e. in the
case devoid of charges and currents.

Section 5 is devoted to find electromagnetic field which can be treated 
as a "plane" wave. The fields are assumed in the form
\begin{equation} F=\psi (\Phi)\,\tilde{F},~~~~~~ G=\psi (\Phi)\,\tilde{G},
\label{wave01}\end{equation}
The factor $\psi $ is a scalar function of the scalar variable $\Phi$,
describing wave-like behaviour, where $\Phi$ represents phase of the wave. 
The solution is found without invoking the wave equation, only Maxwell equations
with the constitutive relation are used.
The $\psi$ contains fast changes in its dependence on $t$ and {\bf r}. 
The second factors $\tilde{F}$ and  $\tilde{G}$ contain much slower dependence.
They are taken as previously found uniform fields.

\section{Magnitudes of the electromagnetic field quantities}

The expression (\ref{fried2}) defines the scalar product of vectors
with the following matrix of the metric tensor $g_{\mu \nu}$:
\begin{equation}g=\left( \begin{array}{cccc}
c^2 & 0 & 0 & 0\\
0 & -a^2 & 0 & 0\\
0 & 0 & -a^2 & 0\\
0 & 0 & 0 & -a^2\\
\end{array} \right) . \label{metric1}\end{equation}

The measure (\ref{fried2}) of distances in
space-time serves to determine separations in proper time $\tau$ for
time-like distances:\footnote{The expression (\ref{fried2})
is divided by $c^2$ in order to obtain the physical dimension of time.}
\begin{equation} d\tau ^2=dt^2-\frac{a^2}{c^2}d{\bf r}^2,
\label{dtau}\end{equation} and distances in proper length for
space-like distances:
\begin{equation} dl^2=a^2d{\bf r}^2-c^2dt^2. \label{dl}\end{equation}
The increments of coordinates are not physically measured
quantities for space or time separations. The metric tensor serves
to calculate them as distances in space-time.
The metric $g$ determines the metric
magnitudes or the lengths for pure space vectors $\Delta {\bf r}$
\begin{equation} ||\Delta {\bf r}||_g=a\sqrt{(\Delta x^1)^2+(\Delta x^2)^2+(\Delta
x^3)^2}. \label{deltax}\end{equation} As follows from (\ref{dtau}),
for pure time vectors the metric magnitude of time separation is
equal to the time separation itself without any additional factor:
\begin{equation} ||d\tau || _g\,=|dt|.\label{dtau2}\end{equation}

\medskip
The expression (\ref{fried2}) defines also the scalar product of
forms with the reciprocal metric tensor $g^{\mu \nu}$ which has the matrix
inverse to (\ref{metric1}):
\begin{equation}g^{-1}=\left( \begin{array}{cccc}
\frac{1}{c^2} & 0 & 0 & 0\\
0 & -\frac{1}{a^2} & 0 & 0\\
0 & 0 & -\frac{1}{a^2} & 0\\
0 & 0 & 0 & -\frac{1}{a^2}\\
\end{array} \right) .\label{metric2}\end{equation}

The electromagnetic field quantities $E,\,D,\,H,\,B$ are exterior
forms, i.e. mappings of line or surface elements into
scalars (for the operational definition of $D$ and $H$ see
\cite{Jancew1}). These mappings determine components of the forms
when space coordinates are given. However, for defining magnitudes
of physical quantities represented by the forms a scalar product is
needed. This is analogous to the line or surface elements: the
components of the named elements can be ascribed for a given
coordinate system, but lengths and areas can be defined only when a
scalar product is introduced.

When scalar product, $g^{-1}$, for one-forms is given, a natural
scalar product, $\hat{g}^{-1}$, for two-forms $K,\,L$ is introduced
through the following formula written for coordinates:
\begin{equation} \hat{g}^{-1}(K,L)=\frac{1}{2}\,K_{\mu \nu}g^{\mu
\alpha}g^{\nu \beta}L_{\alpha \beta}. \label{scaform1}\end{equation}
In particular, the scalar square of the electromagnetic field
strength is
$$\hat{g}^{-1}(F,F)=\frac{1}{2}\left[F_{0j}g^{00}g^{jk}F_{0k}
+F_{k0}g^{kl}g^{00}F_{l0} +F_{kj}g^{kl}g^{jm}F_{lm}\right] .$$
We insert the matrix (\ref{metric2}),
$$\hat{g}^{-1}(F,F)=\frac{1}{2}\,\left[
-F_{0j}\frac{1} {c^2a^2}F_{0j}-F_{k0}\frac{1}{c^2a^2}F_{k0} +\frac{1}{a^4}F_{lj}F_{lj}\right] ,$$
\begin{equation} \hat{g}^{-1}(F,F)=\frac{1}{a^2}\left[
-\frac{1}{c^2}\left( E_1^2+E_2^2+E_3^2\right) +\frac{1}{a^2}\left( B_{12}^2+B_{23}^2+B_{31}^2\right) \right] .
\label{scaform2}\end{equation} Similarly,
\begin{equation} \hat{g}^{-1}(G,G)=\frac{1}{a^2}\left[
-\frac{1}{c^2}\left( H_1^2+H_2^2+H_3^2\right) +\frac{1}{a^2}\left(
D_{12}^2+D_{23}^2+D_{31}^2\right) \right] .
\label{scaform3}\end{equation}

If only the electric part of the field strength is present, its
magnitude, as determined by the metric (\ref{metric2}), is\footnote{The
coefficient $c$ is omitted in order to obtain proper physical
dimension}
\begin{equation} ||E ||_g~=\frac{1}{a}\,\sqrt{E_1^2+E_2^2+E_3^2}.
\label{elmagni}\end{equation}
If the field strength is purely magnetic, its magnitude, determined by $g^{-1}$, is
\begin{equation} ||B||_g~=\frac{1}{a^2}\,\sqrt{B_{12}^2+B_{23}^2+B_{31}^2}.
\label{magmagni}\end{equation}
Similar expressions should be introduced for the electromagnetic
excitations for the magnitudes if pure electric or magnetic
fields exist, respectively:
\begin{equation} ||D||_g~=\frac{1}{a^2}\,\sqrt{D_{12}^2+D_{23}^2+D_{31}^2},
\label{magniel}\end{equation}
\begin{equation}||H||_g~=\frac{1}{a}\,\sqrt{H_1^2+H_2^2+H_3^2}. 
\label{magnimag}\end{equation}

\section{Constitutive relation}

In what follows we use the following notation: $e_0=\frac{\partial}{\partial t}$,
$e_j=\frac{\partial}{\partial x^j}$ are basic vectors, $f^0=\underline{d}t,\,f^i=
\underline{d}x^i$ are basic one-forms, $\underline{d}$ is the exterior derivative
and $\lfloor$ is the contraction.

The constitutive relation between two-forms $F$ and $G$ is
proportional to the Hodge map $\star$ given \cite{Hehl, oziew2} by
the formula\footnote{Also an alternative Hodge map exists
\cite{oziew2} given by $*F=-({\sqrt{-\det g}})^{-1}\,\hat{g}\left(
e_{0123}\lfloor F\right)$ where $e_{0123}=e_0\wedge e_1\wedge
e_2\wedge e_3$ is the volume quadrivector built of the basic vectors.
This formula, however,
leads to the same result for $G=\sqrt{\varepsilon_0/\mu_0}*F$.}
\begin{equation} G=\sqrt{\frac{\varepsilon _0}{\mu_0}}\star F=
-\lambda_0\,\sqrt{-\det g}\, f^{0123}\lfloor \left[ \hat{g}^{-1}
(F)\right] \label{hodge1}\end{equation}
where the bivector in square bracket has the components
\begin{equation} \left[ \hat{g}^{-1}
(F)\right] ^{\mu \nu}=g^{\mu \alpha}g^{\nu \beta}F_{\alpha \beta},
\label{bivect}\end{equation}  and $f^{0123}=f^0\wedge f^1\wedge
f^2\wedge f^3$ is the volume-measure four-form built of the basic
one-forms.

\medskip As the first step in performing the map (\ref{hodge1}) we write the
time-space components of (\ref{bivect})
\begin{equation} \left[ \hat{g}^{-1}(F)\right] ^{0j}=-g^{00}g^{jk}
F_{0k}=-\frac{1}{c^2a^2}\,F_{0j},
\label{tisp}\end{equation} and the space-space components
\begin{equation} \left[ \hat{g}^{-1}(F)\right] ^{jk}=g^{jl}g^{km}
F_{lm}=\frac{1}{a^4} F_{jk}.
\label{spsp}\end{equation} Now we express the bivector in square
bracket of (\ref{hodge1}) by the basic bivectors $e_{\mu
\nu}=e_{\mu}\wedge e_{\nu}$:
$$\hat{g}^{-1}(F)= \frac{1}{a^2}\left[ -\frac{1}{c^2}\left(
F_{01}e_{01}+F_{02}e_{02}+F_{03}e_{03}\right)+\frac{1}{a^2}\left(
F_{12}e_{12}+F_{23}e_{23}+F_{31}e_{31}\right) \right] .$$

The second step in map (\ref{hodge1}) is the contraction with the
basic four-form -- the coefficients do not change, only the basic
bivectors change into complementary basic two-forms: $$f^{0123}\lfloor \left[
\hat{g}^{-1} (F)\right]=\frac{1}{a^2}\left[\frac{1}{c^2}\left( F_{01}f^{23}
+F_{02}f^{31}+F_{03}f^{12}\right)-\frac{1}{a^2}\left( F_{12}f^{03}
+F_{23}f^{01}+F_{31}f^{02}\right) \right] .$$

The last step is multiplication by the numerical factor
$-\sqrt{\frac{\varepsilon _0}{\mu_0}}\,\sqrt{-\det g}=
-\sqrt{\frac{\varepsilon _0}{\mu_0}}\,ca^3$:
$$G=-\sqrt{\frac{\varepsilon
_0}{\mu_0}}\,ca^3\frac{1}{a^2}\left[\frac{1}{c^2}\left( F_{01}f^{23}
+F_{02}f^{31}+F_{03}f^{12}\right) -\frac{1}{a^2}\left( F_{12}f^{03}+F_{23}f^{01}+F_{31}f^{02}\right)
\right]$$
$$=-\sqrt{\frac{\varepsilon_0}{\mu_0}} \left[ \frac{a}{c}\left( F_{01}f^{23}
+F_{02}f^{31}+F_{03}f^{12}\right)-\frac{c}{a}\left(
F_{12}f^{03}+F_{23}f^{01}+F_{31}f^{02}\right) \right] .$$ 
We use
$c=1/\sqrt{\varepsilon_0\mu_0}$:
\begin{equation} G=-\varepsilon _0a\left( F_{01}f^{23} +F_{02}f^{31}+F_{03}f^{12}\right)
+\frac{1}{\mu_0a}\left( F_{12}f^{03}+F_{23}f^{01}+F_{31}f^{02}\right) .
\label{hodge2}\end{equation}
The coefficients in front of appropriate basic two forms are interpreted as components
of the two-form $G=D-H\wedge \underline{d}t$. For instance, the coefficient in front
of the basic two-form $f^{23}$ is $D_{23}$, i.e.
$$D_{23}=-\varepsilon _0aF_{01}=\varepsilon _0aF_{10}=\varepsilon _0aE_1,$$
therefore we obtain:
\begin{equation}D_{23}=\varepsilon_0a\, E_1, ~~~~~D_{31}=\varepsilon_0a\, E_2,
~~~~~D_{12}=\varepsilon_0a\,E_3,\label{de}\end{equation}
\begin{equation} H_1=\frac{1}{\mu_0a}\, B_{23}, ~~~H_2=\frac{1}{\mu_0a}\,
B_{31}, ~~~H_3=\frac{1}{\mu_0a}\, B_{12}.
\label{bh}\end{equation}

One could be prone to treat the coefficients at the right-hand sides
as the permittivity and (the inverse of) permeability of the
vacuum, respectively.\footnote{This can be compared with the relations given
(in different unit system where permeability and permittivity have
the same physical dimension) in \cite{pleb, mash}
$$ \varepsilon _{ij}=\mu _{ij}= (-\det g)^{1/2}\,(g^{ij}/g_{00}), $$ which
for the metric (\ref{metric2}) yield
$\varepsilon_{ij}=\mu_{ij}=a\delta _{ij}$, which is also
proportional to $a$.} The components
$D_{ij},\,B_{ij},\,E_i,\,H_i$, however, are not directly measurable, one
should rather compare magnitudes of the fields.
In this purpose we insert relations (\ref{de}) into (\ref{magniel}):
\begin{equation} ||D||_g~ =\frac{1}{a^2}\sqrt{\varepsilon _0^2a^2\left(
E_1^2+E_2^2+E_3^2\right) }=\frac{\varepsilon_0}{a}\,\sqrt{ E_1^2+E_2^2+E_3^2}
\label{magd}\end{equation}
and use (\ref{elmagni})
\begin{equation}||D||_g\,=\varepsilon_0\,||E||_g.
\label{consted}\end{equation}
We see that the permittivity of the vacuum does not depend on time
and is equal to the electric constant.
We similarly arrive at the relation
\begin{equation} ||B||_g\,=\mu_0\, ||H||_g \label{consthb}\end{equation}
which allows us to claim that the permeability of the vacuum
also does not depend on time and is equal to the magnetic constant.
The product $\varepsilon \mu =\varepsilon_0\mu_0$ does
not depend on time.

\section{Spatially uniform electromagnetic field}

We are going to find field strength $\tilde{F}$ as a simple
two-form, i.e. an exterior product of two one-forms with the
following combinations of basic one-forms $f^{\mu}$:
\begin{equation} k=\xi (t)f^0+k_jf^j,~~~~~h=\zeta (t)f^0+h_if^i,
\label{kl}\end{equation}
where $k_j,h_i=$\, const and summation is present over
repeated indices $i,j=1,2,3$. Thus
$$ \tilde{F}=k\wedge h=\xi (t)f^0\wedge h_if^i+k_jf^j\wedge \zeta (t)f^0+
k_jh_i\,f^{ji}, $$
\begin{equation} \tilde{F}=(\xi h_i-\zeta k_i)f^{0i}+(k_1h_2-k_2h_1)f^{12}+
(k_2h_3-k_3h_2)f^{23}+(k_3h_1-k_1h_3)f^{31}. \label{fkl}\end{equation}
The first term describes the electric field, the three other -- magnetic one.
From the exterior derivative
$$\underline{d}=f^0\frac{\partial }{\partial t}+f^i\frac{\partial }{\partial x^i}$$
only first term gives nonzero contribution on $F$ because no dependence on spatial 
coordinates is present in (\ref{fkl}). In this manner we obtain
\begin{equation}\underline{d}\wedge F=f^0\frac{d}{dt}(\xi h_i-\zeta k_i)\wedge f^{0i}
=(\xi 'h_i-\zeta 'k_i)f^0\wedge f^{0i}=0. \label{maxf}\end{equation}
First Maxwell equation (\ref{flux})is satisfied.

\medskip In order to find the excitation field, we insert the components of $\tilde{F}$
from (\ref{fkl}) into (\ref{hodge2}):
$$\tilde{G}=-\varepsilon _0a\left[ (\xi h_1-\zeta k_1)f^{23}+(\xi h_2-\zeta k_2)f^{31}+
(\xi h_3-\zeta k_3)f^{12}\right] $$
\begin{equation}+\frac{1}{\mu _0a}\left[ (k_1h_2-k_2h_1)f^{03}+
(k_2h_3-k_3h_2)f^{01}+(k_3h_1-k_1h_3)f^{02}\right] . \label{gkl}\end{equation}
We calculate its exterior derivative
$$\underline{d}\wedge \tilde{G}= -\varepsilon _0\{  \,[ (a\xi )'h_1-(a\zeta)'k_1]\, f^{023}
+[ (a\xi )'h_2-(a\zeta)'k_2]\, f^{031}+[ (a\xi )'h_3-(a\zeta)'k_3]\, f^{012}\,\} $$
\begin{equation} +\frac{d}{dt}\left( \frac{1}{\mu _0a}\right) f^0\wedge \left[ (k_1h_2-k_2h_1)f^{03}+
(k_2h_3-k_3h_2)f^{01}+(k_3h_1-k_1h_3)f^{02}\right] . \label{gkl2}\end{equation}
The second term is zero because of exterior product of $f^0$ with $f^{0j}$. In order to ensure
vanishing of the first term we assume $a(t)\zeta (t)=A=$\,const, $a(t)\xi (t)=C=$\,const, that is
\begin{equation} \zeta(t)=\frac{A}{a(t)},~~~~~~\xi(t)=\frac{C}{a(t)}. \label{pafa}\end{equation}
In such a case
\begin{equation} \underline{d}\wedge \tilde{G}=0. \label{g0}\end{equation}
The second (homogeneous) Maxwell equation is satisfied.

\medskip In this manner we have found
\begin{equation}k=\frac{C}{a(t)}f^0+k_jf^j,~~~~~~h=\frac{A}{a(t)}f^0+h_if^i
\label{klAB}\end{equation}
and
\begin{equation} \tilde{F}=k\wedge h =\left[ \frac{C}{a(t)}f^0+k_jf^j\right]\wedge
\left[ \frac{A}{a(t)}f^0+h_jf^j\right] \label{Ffh} \end{equation}
with the following expressions for the components of the uniform electromagnetic field:
\begin{equation} \tilde{F}=\frac{1}{a(t)}(C\,h_i-A\,k_i)f^{0i}+(k_1h_2-k_2h_1)f^{12}+
(k_2h_3-k_3h_2)f^{23}+(k_3h_1-k_1h_3)f^{31}.\label{f0l}\end{equation}

$$ \tilde{G}=-\varepsilon _0\left[ (C\,h_1-A\,k_1)f^{23}+(C\,h_2-A\,k_2)f^{31}+
(C\,h_3-A\,k_3)f^{12}\right] ~~~~~~~~~~~~~~~~~~~~~~$$
\begin{equation}+\frac{1}{\mu _0a(t)}\left[ (k_1h_2-k_2h_1)f^{03}+
(k_2h_3-k_3h_2)f^{01}+(k_3h_1-k_1h_3)f^{02}\right]  \label{gkl3}\end{equation}
with two arbitrary constants $A,\,C$. Since $k_j$ and $h_i$ are constant, the fields 
$\tilde{F},\,\tilde{G}$ do not depend on spatial coordinates, hence the electromagnetic
field is uniform in space. Magnitudes of the electric and magnetic parts are
\begin{equation}||\tilde{E}||_g=\frac{1}{a^2}\sqrt{ (Ch_1-Ak_1)^2+(Ch_2-Ak_2)^2+(Ch_3-Ak_3)^2},
\label{we}\end{equation}
\begin{equation}||\tilde{B}||_g=\frac{1}{a^2}\sqrt{ (k_1h_2-k_2h_1)^2+(k_2h_3-k_3h_2)^2+
(k_3h_1-k_1h_3)^2} \label{wb}\end{equation}
We see that both fields decrease for expanding universe with the same rate.

\medskip In the case of $k_j=0$ only electric field is present:
$$\tilde{F}=\frac{C}{a}\,h_if^{0i}$$
and similarly for $h_j=0$. If one wants the electric field to vanish, one may assume
$Ch_i=Ak_i$, but for $C\neq 0,\,A\neq 0$ the spatial parts of the one-forms $k$ and $h$ are
parallel, so the three last terms in (\ref{f0l}) also are zero and $\tilde{F}=0$ which is
inappropriate. Thus we choose
$A=C=0$, that is, $\xi =\zeta =0$ and then
$$\tilde{F}=k_jf^j\wedge h_if^i$$
expresses the pure magnetic field. Notice that components of this field are constant,
but the magnitude -- according to (\ref{wb}) -- is not.

\section{Plane electromagnetic wave}

An important notion of any wave is its {\it phase}. Let us ponder
what is phase of the plane wave in the Minkowski space-time. It is
the expression
\begin{equation} \Phi (x)= k_{\mu}x^{\mu}=k^{\nu}\eta _{\nu
\mu}x^{\mu}, \label{phase}\end{equation}
where $\eta _{\nu \mu}=\hbox{diag}\{ c^2,-1,-1,-1\}$ is the Minkowski
metric tensor and $k_{\mu }=$\,const.
Which collection of constant numbers: $k_{\mu}$ or $k^{\nu}$ is more
important in this expression? In the language of differential
geometry it is the first one because it establishes components of a
one-form.\footnote{One-form is a linear mapping of the vector space
into scalars, that is into invariants. Minkowski space-time is a
vector space. Therefore it is natural to consider the physical
quantity known as the {\it wave vector} to be a one-form, which
should be rather called {\it wave covector}.} The level surfaces of
the phase, that is, loci of points satisfying $\Phi (x)=$\,const,
are planes and this is the reason why the wave is called plane.

\medskip In more general space-time the phase $\Phi $ needs not be linear function of
coordinates, but still should be a scalar quantity. Its outer
derivative
\begin{equation} k=\underline{d}\Phi=\frac{\partial \Phi}{\partial t}
\,f^0+\frac{\partial \Phi}{\partial x^1}f^1 +\frac{\partial
\Phi}{\partial x^2}f^2+\frac{\partial \Phi}{\partial x^3}f^3,
\label{phase2} \end{equation}
with components $k_{\mu}=\partial _{\mu}\Phi$, is the one-form still called wave covector, but the
components need not be constants. We expect, however, that there
exist coordinates of the space-time in which at least spatial
components $k_i,~i\in\{ 1,2,3\}$ are constant. (Of course,
$k^i=g^{i\mu}k_{\mu}$ are not constant.) For the time-harmonic wave,
the component $\frac{\partial \Phi}{\partial \tau}=k_0=\omega $
should be interpreted as the circular frequency of the plane wave.

\medskip We seek a solution of homogeneous Maxwell equations in the form
\begin{equation} F(t,{\bf r})=\psi (\Phi)\,\tilde{F}(t,{\bf r}),
\label{wave1}\end{equation}
\begin{equation} G(t ,{\bf r})=\psi (\Phi)\,\tilde{G}(t,{\bf r}),
\label{wave2}\end{equation}
where two-forms $\tilde{F}$ and $\tilde{G}$ satisfy homogeneous Maxwell equations:
\begin{equation}\underline{d}\wedge \tilde{F}=0,~~~~~~~ \underline{d}\wedge
\tilde{G}=0. \label{cond}\end{equation}
We take them from previous section as uniform fields.
The factor $\psi $ is a scalar function of the scalar variable $\Phi$,
describing wave-like behaviour. It contains fast changes
in its dependence on $t$ and {\bf r}. The second factors contain
much slower dependence. In the flat space-time,
$\tilde{F}$ and $\tilde{G}$ would be simply constant two-forms.
Usually $\psi $ is taken as a combination of sine and cosine
functions of $\Phi$, which is tantamount to assume that the wave is
time-harmonic. We present our reasoning without this assumption. For
time-harmonic wave, $\tilde{F},~\tilde{G}$ play the role of
amplitudes. The presence of the same function $\psi $ in front of
$\tilde{F}$ and $\tilde{G}$ expresses the synchronicity of changes
of the field strength $F$ and excitation $G$.

\medskip The exterior derivatives of (\ref{wave1}) and (\ref{wave2}) are
$$\underline{d}\wedge F=\psi '(\Phi)\,\underline{d}\Phi \wedge \tilde{F},$$
$$\underline{d}\wedge G=\psi '(\Phi)\,\underline{d}\Phi \wedge \tilde{G}.$$
If the space is devoid of charges and currents, the two Maxwell
equations (\ref{flux}),(\ref{gauss}) are homogeneous and yield
\begin{equation}\underline{d}\Phi \wedge \tilde{F}=0, \label{tilde1}\end{equation}
\begin{equation}\underline{d}\Phi \wedge \tilde{G}=0. \label{tilde2}\end{equation}
Equation (\ref{tilde1}) implies that the two-form $\tilde{F}$ can be factorized 
in the exterior product containing $\underline{d}\Phi$ as one of its factors:
\begin{equation} \tilde{F}=\underline{d}\Phi\wedge h, \label{tilde3}\end{equation}
where $\underline{d}h=0$; the best way to fulfil this condition is
to assume $h_{\mu}=$\,const.  Two-form (\ref{tilde3}) satisfies $\underline{d}\wedge
\tilde{F}=0$, hence the first equation (\ref{cond}) is fulfilled.

\medskip
It would be easy to solve (\ref{tilde2}) by substitution
$\tilde{G}=\underline{d}\Phi \wedge m$ with a one-form $m$ satisfying 
$\underline{d}m=0$, but also the constitutive relation must be satisfied. 
We apply now the map (\ref{hodge1}) (for brevity we introduce the notation $\lambda
=\lambda _0\,\sqrt{-\det g}$~) to (\ref{tilde3}):
$$\tilde{G}=-\lambda \,f^{0123}\lfloor \left[
\hat{g}^{-1}(\tilde{F})\right] =-\lambda \,f^{0123}\lfloor \left[
\hat{g}^{-1}(\underline{d}\Phi\wedge h)\right] 
=-\lambda \,f^{0123}\lfloor \left[ g^{-1}(\underline{d}\Phi)\wedge
g^{-1}(h)\right] .$$
We write this down as
\begin{equation}\tilde{G}=-\lambda \,f^{0123}\lfloor \,(u\wedge v) . ~~~~~~~
~~~~\label{hodge4}\end{equation}  
where we introduced two vectors
\begin{equation} u=g^{-1}(\underline{d}\Phi),~~~~~~ v=g^{-1}(h).
\label{uv}\end{equation}

We have to check whether two-form (\ref{hodge4}) satisfies
condition (\ref{tilde2}). Due to the identity
$$k\wedge \left[ f^{0123}\lfloor \,(u\wedge v)\right]=-f^{0123}\lfloor \,
\{ k\rfloor (u\wedge v)\}$$
valid for any one-form $k$, we substitute $\underline{d}\Phi =k$
and write the condition (\ref{tilde2}) as
\begin{equation} \underline{d}\Phi\wedge \tilde{G}=k\wedge \tilde{G}=
\lambda \,f^{0123}\lfloor \,[k \rfloor (u\wedge v)]=0.
\label{cond2}\end{equation}
Since the contraction with four-form is invertible, the expression in
square bracket must be zero
\begin{equation}k\rfloor (u\wedge v)=k(u)\,v-k(v)\,u=0\label{cond3}\end{equation}
where $k(u)=k_{\mu}u^{\mu}$ is the value of one-form $k$ on vector $u$,
and similarly for $k(v)$. Eq. (\ref{cond3}) indicates that the vectors $u$ and
$v$ are parallel for nonzero scalars $k(u)$ and $k(v)$, and
this along with (\ref{hodge4}) would imply that $\tilde{G}=0$ which
is undesirable. Thus the equalities
\begin{equation}\underline{d}\Phi(u)=k_{\mu}u^{\mu}=0,~~~~~~~~~
\underline{d}\Phi(v)=k_{\nu}v^{\nu}=0\label{cond4}\end{equation}
are necessary to satisfy condition (\ref{cond2}) and, therefore, (\ref{tilde2}).
Obviously, they are also sufficient. We know from (\ref{uv}) that $u=g^{-1}(k)$,
i.e. $u^{\mu}=g^{\mu \alpha}k_{\alpha}$, similarly $v^{\nu}=g^{\nu
\beta}h_{\beta}$, so the conditions (\ref{cond4}) are
$k_{\mu}g^{\mu \alpha}k_{\alpha}=0$, $k_{\nu}g^{\nu \beta}h_{\beta }=0$,
or
\begin{equation} g^{-1}(\underline{d}\Phi,\,\underline{d}\Phi)=0,
~~~~~~~ g^{-1}(\underline{d}\Phi,\,h)=0.
\label{orto}\end{equation} 
Thus we conclude that equation
(\ref{tilde2}) is satisfied if and only if the one-forms
$\underline{d}\Phi,\,h$ from factorization (\ref{tilde3}) satisfy
conditions (\ref{orto}). The first one says that the wave covector
$k$ is orthogonal to itself. Let us look at this closely.

\medskip We apply the metric tensor (\ref{metric2}) to the one-form
(\ref{phase2}):
$$ \frac{\omega ^2}{c^2}-\frac{1}{a^2}\left( k_1^{~2}+k_2^{~2}+k_3^{~2}
\right) =0,$$ hence
\begin{equation} \omega ^2=\frac{c^2}{a^2}\left(
k_1^{~2}+k_2^{~2}+k_3^{~2}\right) \label{czest0}\end{equation}
and
\begin{equation}\omega =\pm \frac{c}{a}\, \sqrt{k_1^{~2}+k_2^{~2}+k_3^{~2}}
=\pm \frac{c}{a}\,|{\bf k}|. \label{omega1}\end{equation}
We see that $k_i$ can be constants, but $\omega $ can not. By comparing
(\ref{omega1}) with (\ref{phase2}) we obtain
$$\frac{\partial \Phi}{\partial x^i}=k_i, ~~~~\frac{\partial \Phi}{\partial
t}=\pm \frac{c}{a(t)}|{\bf k}|.$$
After introducing the indefinite integral
\begin{equation} b(t)=\int \frac{c}{a(t)}\,dt \label{int}\end{equation}
we are allowed to write down
\begin{equation} \Phi (x)=\pm \,b(t)|{\bf k}|+k_ix^i \label{phix}\end{equation}
as a phase of the plane electromagnetic wave.
By taking $h$ as purely spatial one-form orthogonal to $k$ we obtain
\begin{equation} \tilde{F}=\left[ \pm \frac{c}{a(t)}|{\bf k}|\,f^0+k_1f^1+k_2f^2+k_3f^3\right]\wedge h.
\label{ftild}\end{equation} which is particular uniform electromagnetic field
(\ref{Ffh}), with $A=0$, found in previous section.

\medskip
The substitution of the phase (\ref{phix}) into (\ref{tilde3})
and (\ref{wave1}) yields the explicit field strength of the plane
electromagnetic wave:
\begin{equation} F(t,{\bf r})=\psi \left( \pm b(t)\,|{\bf k}|+k_ix^i\right)
\left[ \pm \frac{c}{a(t)}|{\bf k}|\,f^0+k_if^i\right]\wedge h_jf^j. 
\label{wave4}\end{equation}
The electric part of it, according to (\ref{faradsplit}) is determined by
the first term in square backet:
$$E\wedge \underline{d}t=E\wedge f^0=\pm \frac{c}{a(t)}\,\psi \,|{\bf k}|\,f^0\wedge h,$$
hence the electric field one-form is
\begin{equation} E(t,{\bf r})=\mp \frac{c\psi (t,{\bf
 r})}{a(t)}\,|{\bf k}|\,h \label{e-field}\end{equation}
The magnetic part of (\ref{wave4}) is the magnetic induction
two-form
\begin{equation} B(t,{\bf r})=\psi (t,{\bf r})\, [(k_1h_2-k_2h_1)\,
f^{12}+(k_2h_3-k_3h_2)\, f^{23}+(k_3h_1-k_1h_3)\, f^{31}]
\label{b-field}\end{equation}

To write them down in terms of traditional vectors we simly
equate the one-form components $E_i$ to vector components
$E^i$:
$$E^i=E_i $$
and the two-form components $B_{ij}$ to pseudovector components
$B^l$:
$$B^1=B_{23}, ~~ B^2=B_{32}, ~~ B^3=B_{12}.$$
Then the expressions (\ref{e-field}, \ref{b-field}) assume 
the form
\begin{equation} \vec{\bf E}(t,{\bf r})=\mp \frac{c\psi (t,{\bf
		r})}{a(t)}\,|{\bf k}|\,\vec{\bf h}, \label{e-fieldv}\end{equation}
\begin{equation} \vec{\bf B}(t,{\bf r})=\psi (t,{\bf r})\, 
\vec{\bf k}\times \vec{\bf h}.
	\label{b-fieldv}\end{equation}

Formula (\ref{e-fieldv}) describes a linearly polarized wave with coordinate dependence $\psi (t,{\bf r})a(t)^{-1}$ and
amplitude parallel to $\vec{\bf h}$, establishing the polarization direction. The scale factor $a$ is present only in the electric
part, but when calculating magnitudes according to
(\ref{elmagni}, \ref{magmagni}), the scale frctor is appears
with the same power in both fields:
$$||E||_g=\frac{c|\psi |}{a^2}\,|{\bf k}|\,|{\bf h}|,~~~~
||B||_g=\frac{|\psi |}{a^2}\, |{\bf k}|\, |{\bf h}|. $$

\medskip For the sake of completeness, we shall find 
the two-form $G$. We repeat here formula (\ref{hodge2}):
\begin{equation} G=-\varepsilon _0a\left( F_{01}f^{23} +F_{02}f^{31}+F_{03}f^{12}\right)
+\frac{1}{\mu_0a}\left( F_{12}f^{03}+F_{23}f^{01}+F_{31}f^{02}\right) .
\label{gaus1}\end{equation}
We find from (\ref{wave4})
$$ F_{0j}=\pm \psi\,\frac{c|{\bf k}|}{a}\,h_j,~~~~~~F_{ij}=\psi \, (k_ih_j-k_jh_i) $$
and insert into (\ref{gaus1})
$$G(t,{\bf r})=\psi \left( \pm b(t)\,|{\bf k}|+k_ix^i\right)\,
\{  \mp \sqrt{\varepsilon _0/\mu_o}\, |{\bf k}|\,(h_1f^{23}+h_2f^{31}+h_3f^{12}) 
~~~~~~~~~~~~~~~~~~~~~~$$
\begin{equation}~~~~~~~~-\frac{1}{\mu _0a(t)}\, 
[(k_1h_2-k_2h_1)\, f^3+(k_2h_3-k_3h_2)\, f^1+(k_3h_1-k_1h_3)\, 
f^2]\wedge f^0 \}.\label{gaus2}\end{equation}
The second term in curly bracket determines the magnetic field one-form
\begin{equation} H(t,{\bf r})=-\frac{\psi(t,{\bf r})}{\mu _0a(t)}\,
[(k_1h_2-k_2h_1)\, f^3+(k_2h_3-k_3h_2)\, f^1+(k_3h_1-k_1h_3)\, 
f^2], \label{h-field}\end{equation}
whereas the first term is connected with the electric 
induction two-form
\begin{equation} D(t,{\bf r})=\mp \psi(t,{\bf r})\,\sqrt{\varepsilon 
_0/\mu_0}\, |{\bf k}|\,(h_1f^{23}+h_2f^{31}+h_3f^{12}) 
\label{d-field}\end{equation}

\section{Magnitudes of some observable quantities}

We now consider the measurable quantities of the phase, 
i.e. the magnitudes
determined by the metric (\ref{metric2}):\footnote{The
 coefficient $c$ is omitted after first equality in 
 order to obtain appropriate physical dimension}
\begin{equation} ||\omega ||_g\,=|\,\omega|=
\frac{c}{a(t)}\,\sqrt{k_1^{~2}+k_2^{~2}+k_3^{~2}},
\label{omega2}\end{equation}
where $|\omega |$ is the absolute value of the real number $\omega $, and
\begin{equation} ||{\bf k}||_g\,=\frac{1}{a(t)}\sqrt{k_1^{~2}+k_2^{~2}+k_3^{~2}}.
\label{falist}\end{equation} We see that the measured circular
frequency and the measured magnitude of the wave covector decrease
with time when the Universe expands. This corresponds to the
observation that the light from distant galaxies is shifted to the
red end of spectrum which is called the redshift. By comparing the
magnitudes (\ref{omega2}) and (\ref{falist}) we obtain
\begin{equation}||\omega ||_g=c\,||{\bf k}|| _g,
\label{czest2}\end{equation} 
which means that the phase velocity
$||\omega ||_g/||{\bf k}||_g$ of the
plane wave is constant in time and equals $c$.

\vspace{4mm}
The energy density of the electromagnetic field is given
by the formula
\begin{equation} w=\frac{1}{2}\,(E\wedge D+H\wedge b).
\label{w1}\end{equation}
Hence, we calculate the needed products
$$ E\wedge D=\frac{\psi ^2c\,|{\bf k}|^2|{\bf h}|^2}{\mu_0\,a(t)}\,f^{123}, $$
$$ H\wedge B=\frac{\psi ^2c\,|{\bf k\times  h}|^2}{\mu_0\,a(t)}\,f^{123} $$
where $f^{123}=f^1\wedge f^2\wedge f^3$ is the basic three-form.
We choose ${\bf k\perp h}$, then
$$H\wedge B=\frac{\psi ^2c\,|{\bf k}|^2|{\bf h}|^2}{\mu_0a(t)}\,f^{123}=E\wedge D,$$
so the energy density three-form is
\begin{equation}w=\frac{\psi ^2c\,|{\bf k}|^2|{\bf h}|^2}
{\mu_0\,a(t)}\,f^{123}. \label{w2}\end{equation}
A magnitude of the spatial three-form is its coordinate
in front of $f^{123}$, multiplied
by $\sqrt{-g^{11}\,g^{22}\,g^{33}}=\sqrt{a^{-6}}=a^{-3}$
which is
\begin{equation} ||w||_g=\frac{\psi ^2c\,|{\bf k}|^2
|{\bf h}|^2}{\mu_0\,a(t)^4}. \label{w4}
\end{equation}

How can one interprete factor $a^4$ in the denominator? 
Let us look at the electromagnetic wave as a collection
of photons. Let in a cube with edge $l$ be $N$ photons
in time $t_0$ [when $a(t_0)=1$], so the concentration of photons is
$n_0=N/l^3$. In another time $t$ of the Universe
the cube has its edge with length $a(t)l$, hence the
volume is $a(t)^3l^3$. Therefore, the concentration of
photons changes with time:
\begin{equation} n(t)=\frac{n_0}{a(t)^3}.
\label{nt}\end{equation}
Each photon has its energy propotional to $||\omega ||$:
$\varepsilon =\hslash \,||\omega ||$.
According to (\ref{omega2}) $||\omega (t)||_g=
\frac{\omega _0}{a(t)}$, so the energy of single photon also
changes with time:
\begin{equation} \varepsilon (t)=\frac{\hslash \omega _0}{a(t)} .
\label{om0}\end{equation}
The energy density of the photons is the product of 
(\ref{nt}) and (\ref{om0}):
\begin{equation} ||w(t)||=\hslash \,\frac{n_0}{a(t)^3}\,
\frac{\omega _0}{a(t)}=\hslash \frac{n_0\omega _0}{a(t)^4}
\label{wt}\end{equation}
Now the factor $a^{-4}$ in (\ref{w4}) becomes
natural.

\section{Conclusion}

We have considered the electromagnetic field in expanding universe
described by the spatially flat Friedman model. We take the field 
to be weak, which means that its energy momentum tensor does not 
influence the gravitational field.

The scalar product matrices ($g$ for vectors, $g^{-1}$ for
one-forms) govern not only distances in space-time but also
magnitudes of other physical quantities. We have presented the
magnitudes of electromagnetic field quantities $E,\,D,\,B,\,H$ --
the scale factor is present there. The matrix $g^{-1}$
enters the constitutive relation between $F$ and $G$ which implies
that the scale factor is present in the relations
between components of $D$ and $E$ on the one hand and between those
of $B$ and $H$ on the other. But after comparing the magnitudes of
these quantities it turns out that the permittivity and the
permeability of the vacuum are constant and are the same as
in Maxwell-Lorentz electrodynamics.

The time variation of the permittivity and permeability of the
vacuum was concluded in the literature from relations not between
magnitudes but between components of $D$ and $E$  and those of $B$
and $H$ \cite{pleb, mash, sumner}. One of the authors \cite{sumner}
even speculated on the time variation of the fine structure constant
in which permittivity is present. In our opinion this is
superfluous.

First aim of the paper is obtaining explicit solution of
the homogeneous Maxwell's equations in the form of uniform field
allowing its dependence only on time. It turns out that for the scale factor $a$
magnitudes of both electric and magnetic parts of the field depend on time as $1/a$.
[Compare eqs. (\ref{elmagni}) and (\ref{magd}).]

The main task is finding explicit solution of the
Maxwell's equations in the form of a plane electromagnetic wave
without invoking the wave equation. It was claimed in
\cite{pleb,mash} that equations of electromagnetic field in
presence of gravity can be interpreted as the Maxwell equations in
flat space-time but in a medium characterized by permittivities
depending on metric coefficients. In the present approach,
the Maxwell equations are metric independent and the metric
enters only the constitutive relations. It turned out that the 
permittivities are the same as in the flat Minkowski space-time.

In the literature about electromagnetic waves in gravitational
fields, two approaches occur. In first one (see \cite{pleb}), the
light rays are considered and the fields themselves are found in a
kind of ``Born approximation". Our solution is presented without any
approximation. In second approach (see \cite{mash, yao, haghi}), a
time-harmonic solutions with a definite parity and angular momentum
are found. Of course, time-harmonic plane wave can be represented in
the basis of spherical harmonics by the Gegenbauer expansion. This
is, however quite long way: (i) find the spherical waves in not so
short derivation, (ii) find the limit of the Gegenbauer series.

Our proposed way is not limited to time-harmonic waves and is
shorter: consider fields in the form (\ref{wave1},\ref{wave2}) with
synchronous dependence $\psi$ on the phase function $\Phi$. The factors
$\tilde{F},\,\tilde{G}$ are slowly changing fields which can be treated
as amplitudes of the wave. Our
assumption is that the phase of the wave has constant spatial components
of the wave covector, so we are allowed to call it a plane wave. The
Maxwell equations lead to the conditions (\ref{tilde1},\ref{tilde2})
for the amplitudes $\tilde{F},\,\tilde{G}$. First equation can be
solved by substitution (\ref{tilde3}), the second one, combined with
the constitutive equation, implies the conditions (\ref{orto}). The
first of them imposes a relation between time and space components
of the wave covector $k$. The plane wave field strength is
$$F(\tau ,{\bf r})=\psi (\Phi)\,k\wedge h,$$ where the phase $\Phi$ is
expressed by (\ref{phix}) and $h$ is the polarization one-form.

\medskip The magnitudes of the circular frequency and of the wave covector 
depend only on time, decreasing by the factor $a^{-1}$. Since both 
decrease by the same factor, the phase velocity of the wave is 
constant and equal to $c$.
In the literature about electromagnetic waves in expanding 
universe only Mashhoon \cite{mash} has shown classically that 
the frequency of the wave depends on the 
metric.\footnote{Predominantly, the	explanation of the decrease 
of frequency is done by passing to the quantum picture, namely by 
considering a photon moving through	space-time endowed with the 
Friedman metric, see \cite{schutz}, Sec	12.6. The photon loses 
its energy $E$ by the factor $a^{-1}$ hence, by the Planck 
relation $E=\hbar \omega$, the same concerns its frequency.}
The energy density of the electromagnetic wave changes in
time with the factor $a_4$, which can be confirmed also by 
consideration of the concentration of photons and the density
of their energy.

\medskip The decreasing of frequency is connected with the observed redshift
of light arriving from distant galaxies. People not working in
cosmology think that galaxies run away and the redshift is a result
of the Doppler effect. The Friedman metric, however, is derived
under assumption that the matter (i.e. the galaxies) rests in the
chosen coordinate frame. In other words, the coordinates, in which
the scale factor $a$ is a function of time only, are
distinguished by the fact that the galaxies do not move. The
distance between them grows because the space is expanding. Our
result (\ref{omega2}), (\ref{falist}) shows that the redshift is
only a manifestation of the expansion. One may ask the question: why
the light is shifted to the red if the galaxies do not move? The
answer is: because the light from distant objects travels very long
in time, the scale factor increases  during the travel and the light
frequency is diminished by this factor.

\medskip One could ponder on question whether similar reasoning
can be performed for general metric (\ref{fried1}) of the Friedman model. In such a case the metric tensor is not so
simple as in eq. (\ref{metric1}), hence in the constitutive
equation (\ref{hodge2}), in addition to the time-dependent
factor $a$, also another space-dependent factors must be
present. Therefore, the exterior derivative (\ref{gkl2})
would be much more complicated and it is not sure whether spatially uniform electromagnetic field exists in this
situation.

\end{document}